\documentclass[epj]{svjour}
\usepackage{graphicx}
\usepackage{amssymb,amsmath}
\begin{document}
\title{{\large Topical Issue on:}\\
Frontiers in Nuclear, Heavy Ion and Strong  Field Physics}
%\subtitle{Dedicated to Walter Greiner; October 1935 -- October 2016}
\titlerunning{Frontiers in Nuclear, Heavy Ion and Strong  Field Physics}
\author{Tamas Bir\'o\inst{1}% 
\and Carsten Greiner\inst{2}% etc
\and Berndt M\"uller\inst{3}% etc
\and Johann Rafelski \inst{4}% etc
\and Horst St\"ocker\inst{5}% etc 
}   % Do not remove
%
%\offprints{}  % Insert a name or remove this line
%
\institute{Institute for Particle and Nuclear Physics, Wigner RCP, Budapest, Hungary \and
Institut f\"ur Theoretische Physik der Goethe-Universit\"at Frankfurt, Germany \and
Department of Physics, Duke University, Durham, NC 27708-0305, USA \and
Department of Physics, The University of Arizona, Tucson, Arizona, 85721, USA \and 
GSI Darmstadt and FIAS  Goethe-Universit\"at Frankfurt, Germany}
\date{January  2018}
\large
\abstract{In memoriam: Walter Greiner: 29 October 1935 -- 5 October 2016\\[1cm]%
%%%%%%%%%%%%%%%%%%%%%%%%%%%%
%\begin{figure}
\centerline{\hspace*{-2cm}\includegraphics[width=1.6\columnwidth]{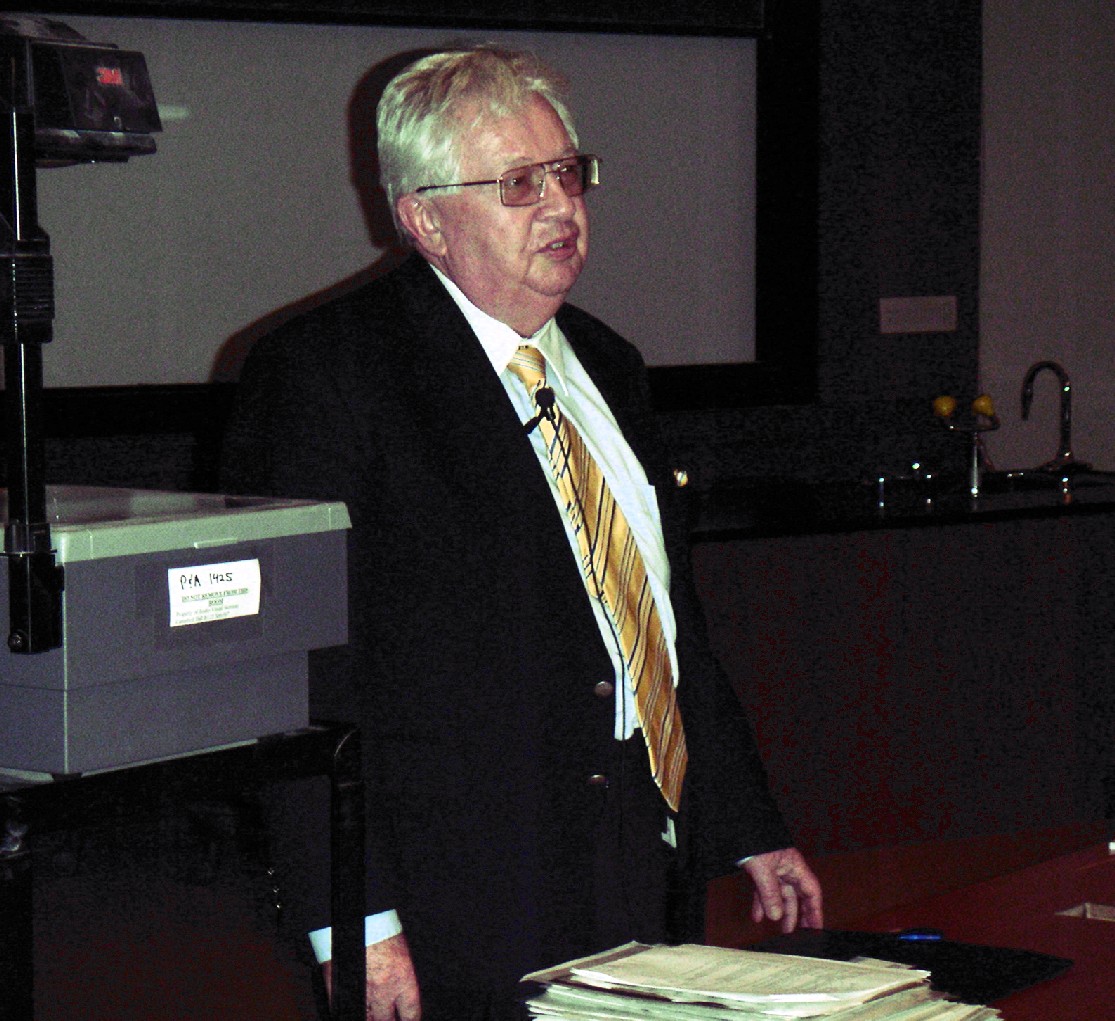}}
%\caption{
Walter Greiner lecturing on March 31, 2006 at the Strangeness in Quark Matter (SQM06) conference at the University of California, Los Angeles {\it Photo by J. Rafelski}.
%}\label{WG06}
%\end{figure}
%%%%%%%%%%%%%%%%%%%%%%%%%%%%%%%%%%%%%%%%%%%%%%%%%%%%%%%%%%%%%%%%%%%%
\PACS{%
20. NUCLEAR PHYSICS 
 } % end of PACS codes
 } %end of abstract
\maketitle\onecolumn
%%%%%%%%%%%%%%%%%%%%%%%%%%%%%%%%%%%%%%%%%%%%%%%%%%%%%%%%%%%%%%%%%%%%%%% 
A special issue of the {\em European Physical Journal A} (February 2018, in press) is dedicated to one of our most distinguished colleagues,‎ Walter Greiner, internationally renowned scholar, creative thinker, scientific pioneer, our teacher and friend. We focus here on Walter as eminent researcher, educator, mentor of young scientists, and founder of scientific institutions.

Professor Walter Greiner: Director of the Institut f\"ur Theoretische Physik at the Goethe Universit\"at in Frankfurt am Main from 1965 until 1995, and  until his death  Professor Emeritus; and Founding Director of, and Senior Fellow at FIAS, the Frankfurt Institute for Advanced Studies, from 2004-2016; is recognized worldwide for his pioneering role in establishing heavy-ion physics as an independent, rapidly growing research field, and as one of the founders of the Gesellschaft f\"ur Schwerionenforschung (GSI), the Helmholtz Center for Heavy-Ion Research, as well as of the multidisciplinary Frankfurt Institute for Advanced Studies (FIAS).

Walter\lq s interest in science was phenomenal and not limited to a particular field. His creativity was astounding: he contributed to more than 1000 published articles, in a wide variety of fields. These scientific achievements earned him eleven honorary doctorates from renowned universities all over the world. He won the Max-Born Prize, jointly given by the Institute of Physics, London, and the German Physical Society, the Otto-Hahn Prize of the City of Frankfurt am Main, the Alexander von Humboldt Medal, and he was an Officier dans l\rq Ordre Palmes Acad\'emiques.

Walter earned his Dr. rer. nat. (Ph.D.) with a dissertation on nuclear polarizability in muonic atoms, under the guidance of his mentor, Hans Marschall, at the University of Freiburg in 1961. From 1962 to 1964 he was assistant professor at the University of Maryland, were he met his lifelong friend and colleague, then at the University of Virginia, Judah Eisenberg, the coauthor of the seminal three-volume monograph \lq\lq Theoretical Nuclear Physics\rq\rq, an indispensable textbook for generations of graduate students of nuclear physics.

Upon his arrival in Frankfurt in 1965, Walter reshaped the physics curriculum by introducing a core course sequence in theoretical physics for students starting in their first semester. The course was offered for the first time in 1968 when several of us were in the classroom. This was the beginning of the \lq\lq Frankfurt School\rq\rq of Theoretical Physics at the J.W. Goethe University, which quickly began to attract students from all over the world to pursue world-leading research in nuclear and heavy-ion physics. Under the tutelage of Walter Greiner, more than 150 students earned their doctorates. More than 40 of these became leading scholars and researchers of theoretical physics and heavy-ion science worldwide.
 
Looking back at 1968, one recognizes that there was no textbook geared to Walter\rq s level of theory instruction for entering students. Walter addressed this problem through the formation of student groups charged with preparing write-ups of the material covered in his lectures, which he selected from a multitude of sources. Within a decade these notes turned into a renowned series of theoretical physics textbooks, unique in their scope, starting at the entering student level but leading into advanced topics -- today these lecture notes on Theoretical Physics are published in a series consisting of more than 15 volumes and have been translated into several languages. They have taught generations of physics students the foundations of theoretical physics, and they are highly appreciated as a reference source due to the richness of their content.

However, Walter\rq s pioneering spirit did not focus on education alone. Already in 1966 he started an initiative to create a national heavy-ion research laboratory and invited his colleagues at the Hessian universities to participate. A joint proposal was submitted to the Federal Ministry of Research and Technology and to the State of Hesse, which was approved in 1969. The new laboratory, with the heavy-ion accelerator UNILAC at its center, was completed in 1975 and evolved into today's world-leading laboratory for heavy-ion research, known as GSI, and presently expanded into the international Facility for Antiproton and Ion Research (FAIR).

The theoretical basis for this new field of research had already been established by Walter together with collaborators and students recruited by the 1968 teaching initiative, appropriately called the Frankfurt School of Theoretical Physics. Three general scientific topics became the signature projects of Walter and his circle of students and collaborators:
\begin{itemize}
\item Walter\rq s interest in collective properties of nuclei and of nuclear fission was extended to super-heavy nuclei, initiating an international race for the discovery of new elements that has now reached element 118.
\item Strong-field physics, culminating in the phenomenon of positron production by vacuum decay, a project that two of us (Berndt and Johann) were deeply involved in. This laid foundation for the new research field \lq\lq QED of Strong Fields\rq\rq\ which is today motivating two diverse fields of physics: ultra intense laser-matter interactions and (peripheral) ultra-relativistic heavy ion collisions. 
\item In 1974 together with one of us (Horst) Walter became actively involved in relativistic heavy ion science, predicting the occurrence of shock waves and the collective flow of dense and hot strongly interacting matter.
\end{itemize}
The tight links of these activities to the properties of nuclear equation of state, exposed in high energy heavy ion collision laboratory experiments around the world, from the early days in Berkeley and Dubna, to GSI/FAiR, Brookhaven-RHIC and CERN-LHC, and to cosmology and astrophysics tying the very small to the very large are evident. The novel field of Relativistic Heavy Ion Physics now comprising thousands of scientists worldwide studying the phase structure of dense and hot nuclear matter and  hot QCD matter, the quark-gluon plasma was indeed born at GSI and in Frankfurt.

Walter was not only driven by a life-long passion for physics, he was also profoundly concerned with the impact of science on society. He realized that scientific collaborations can serve as means of reconciliation. Thus, he fostered collaboration with scientists in the United States of America, where Michael Danos at the National Bureau of Standards in Washington, DC (now NIST Gaithersburg) became his close life-long collaborator and friend. 

He also made good use of the Alexander von Humboldt Foundation's Senior U.S. Scientist Award program to bring many Jewish scientists who had to flee prosecution in Germany in their youth during the Nazi era, such as Walter Meyerhof and Eugen Merzbacher, back to Germany for the first time for extended visits. In the same spirit together with Judah Eisenberg, who was now teaching in Tel-Aviv, Israel, he initiated a fruitful scientific exchange between Israeli and German universities. Walter hosted all in all more than one hundred Humboldt fellows from all over the world.

\begin{figure}
\centerline{%
\includegraphics[width=0.8\columnwidth]{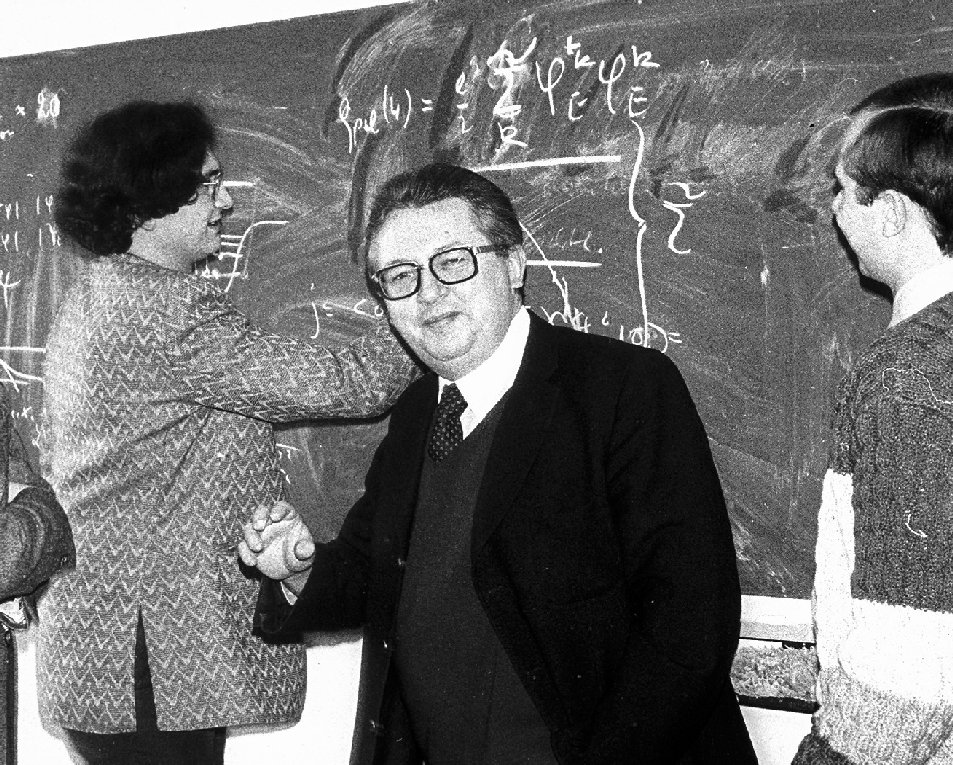}
}
\caption{Strong fields working group met each Saturday in early 70s. The picture shows Walter Greiner, with Johann writing and Berndt watching. {\it J. Rafelski photo archives}}
\label{WG70}
\end{figure}

Walter loved to work with students. Everybody remembers the tea cart that accompanied the palaver meetings on Friday afternoon when students reported their progress. Less known are the  Saturday morning strong field group meetings. Figure\,\ref{WG70} shows how things worked. Walter also invited his students to accompany him in his frequent research visits in USA, and two of the contributing authors now work in United States. 

%%%%%%%%%%%%%%%%%%%%%%%%%%%%
%%%%%%%%%%%%%%%%%%%%%%%%%%%%

In 1970 Walter invited  Johann to accompany him on a trip to work at NBS (NIST today). Among our activities was a seminar visit to Stony Brook on Long Island. From there Walter took, as a last minute development, a ferry to Yale where he met D. Allan Bromley, who later shaped science policy for President George H.W. Bush. This encounter was a seminal development for heavy-ion physics, creating not only a long and deep friendship, but also initiating a coordinated US-German push into heavy-ion physics in low, medium and later high energy domains, without which RHIC may not have happened.

Walter was also acutely aware of the fact that fundamental science must reach society in order to receive wide acceptance and recognition. Together with prominent citizens of Frankfurt, he founded the \lq\lq F\"orderverein f\"ur Physikalische Grundlagenforschung\rq\rq\, which supports young and senior scientists at Goethe University and FIAS. Here, very much in the tradition of Goethe-Universit\"at as an institution founded by the citizens of Frankfurt am Main, sponsors generously donate funds that are used to award students and members of the Department of Physics and of FIAS for their achievements in research and teaching. This initiative was instrumental in the creation of the Frankfurt Institute for Advanced Studies, FIAS, with its interdisciplinary theoretical science mission, encompassing a wide spectrum of disciplines ranging from life science to physics.

In these many professional activities showing restless devotion to physics and science, Walter never neglected family or himself.  He found time for his personal recreation at his home in the Taunus mountains, or at his vacation house in Spain doing swimming, hiking and picking mushrooms. He had the true privilege of being married for more than 55 years, and greatly benefited from the capable support by his beloved wife B\"arbel who took up the complete private life management. Together they have raised two sons, both physicists, who  also both became professors, and in recent years he enjoyed watching the five grandchildren grow. Walter also loved to travel. With his wife he found time for visiting and getting to know exotic places around the world.

Walter was a tireless leader in his personal and professional life. He cared intensely about all those close to him, family, students and friends. He was deeply concerned about the health of Nuclear Science and its impact on society. He was a highly respected mentor for his students, and a warm friend for his colleagues, who were many, from all over the world, representing his passion for the global nature of science.

We are certain that his legacy of contributions to our field and the worldwide collaborative interactions he helped form will continue. In this spirit this special volume comprising 18 high quality and fully refereed research papers is published today by the {\em European Physical Journal A} as a testimonial to Walter Greiner\rq s lasting scientific influence.

%%%%%%%%%%%%%%%
\end{document}